\documentclass[doublecol]{epl2}
\usepackage{graphicx,bm,amsmath,amssymb}
\makeatletter
\let\epl@pacsmissing\relax
\makeatother

\title{Spinor bosons realization of the SU(3) Haldane phase with adjoint representation}
\shorttitle{Spinor bosons realization of the SU(3) Haldane phase with adjoint representation}
\author{Junjun Xu \thanks{E-mail: \email{jxu@ustb.edu.cn} (corresponding author)}}
\shortauthor{Junjun Xu}

\institute{Beijing Weak Magnetic Detection and Applied Engineering Technology Research Center, School of Mathematics and Physics, and Institute of Theoretical Physics, University of Science and Technology Beijing, Beijing 100083, China}

\abstract{The Haldane phase with local SU(3) adjoint representation constitutes the simplest non-trivial symmetry-protected topological phases in the SU($N>2$) Heisenberg spin chains. In this paper, we propose to realize this phase by a two-species spinor Bose gas, with each species labeling the quark or antiquark states of SU(3) symmetry. In the strong-coupling limit, we determine the ground-state phase diagram, and identify a quantum phase transition from the Haldane phase to a dimer phase. We show how to characterize the Haldane phase through its edge excitations. We also explain the physics at the dimer phase, by constructing an explicit ground-state ansatz at the dimer point.}

\begin{document}

\maketitle

\section{Introduction}
The Haldane conjecture, which predicts a gapped ground state in antiferromagnetic Heisenberg (AFH) chains with local SU(2) integer-spin representations \cite{Haldane83, Affleck87, Affleck89}, has reshaped our understanding and classification of quantum matters, and leads to the discovery of the symmetry-protected topological (SPT) phases \cite{Wen, Pollmann10, Pollmann12}.
Recently, theoretical progress in this direction generalizes the Haldane conjecture to SU($N$) AFH spin chains, and finds a gapless ground state when the number of boxes in the Young diagram $p$ is coprime with $N$, which is connected to the ${\rm SU}(N)_1$ Wess-Zumino-Witten conformal field theory, while for other representations field theory calculations predict a finite mass gap \cite{Affleck17, Oshikawa19, Affleck20}. Interpretations about this mass gap include the existence of spinon confinement \cite{Greiter07a, Greiter07b, Greiter09}, and more recently the fractional topological excitations called instantons \cite{Wamer20}.

Compared to the SU(2) case, the richer representations of the SU($N$) symmetry are predicted to support new features in their gapped states. For example, the SPT phases in SU($N$) AFH spin chains can be protected by the $Z_N\times Z_N$ symmetry, and have $N-1$ distinct non-trivial classes  \cite{Quella12, Else}, corresponding to different elements of the second cohomology group $H^2(Z_N\times Z_N, U(1))=\mathbb{Z}_N$, which can be revealed by the representations of different edge states \cite{Furusaki14, Affleck22}. At the same time, the ground state of the SU($N$) system can have degeneracies if these non-trivial classes are symmetrically connected.

On the numerical front, the minimal model to check above new features is to consider the SU(3) symmetry.
For the symmetric representation with Young diagram $[3~0~0]$, large-scale density matrix renormalization group (DMRG) results have recently determined the ground state energy gap to be around $0.04J$ \cite{Mila20}. However, this SPT state is found to be trivial \cite{Lecheminant20}, which is confirmed more recently by tensor-network calculations \cite{Verstraete}.
For non-symmetric representations, recent Monte Carlo simulations together with field theory calculations show there is always a gapped phase in the SU(3) self-conjugate representations \cite{Wamer19}, where different from the SU(2) case, the flavor-wave calculation shows six Goldstone bosons with two different velocities, and the topological angle becomes non-trivial with a spontaneously broken parity symmetry for an odd number of boxes.

\begin{figure}[h]
    \centering
    \includegraphics[width=0.45\textwidth]{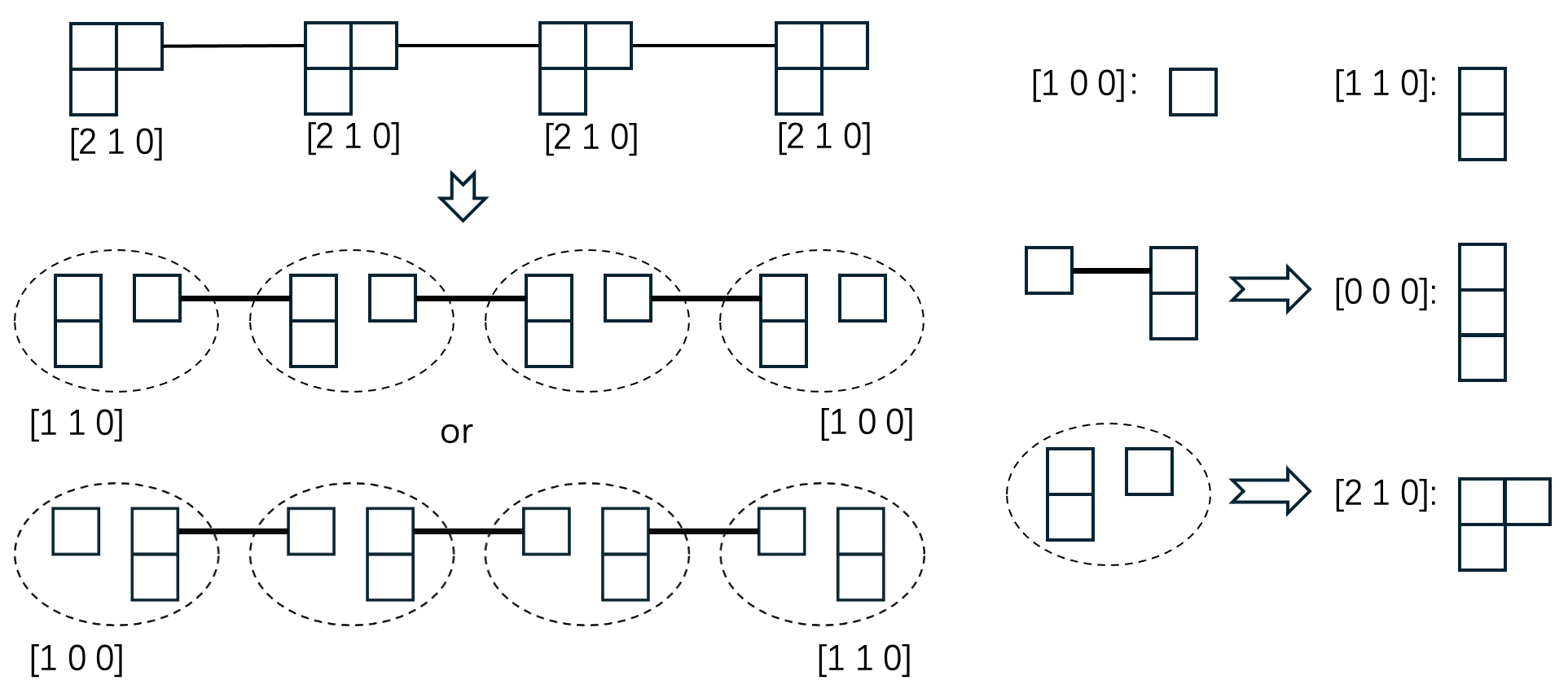}
    \caption{An illustration of the AKLT valence-bond states with local SU(3) adjoint representation [2~1~0] shown as their Young diagrams, where the solid lines map two neighbour virtual states to a SU(3) singlet, and the dashed circles map the local virtual states to a physical octet state. The appearance of two inequivalent virtual states, as shown in dashed circles, gives rise to two distinct chiral topological states with different edge modes.}
    \label{fig:aklt}
\end{figure}

Thus, the simplest representation in SU(3) symmetry with non-trivial topological properties would be the adjoint representation $[2~1~0]$. In this system, the edge states of the two sides can be in the representation $[1~0~0]-[1~1~0]$ or $[1~1~0]-[1~0~0]$, corresponding to the two classes of SPT phases, where each class breaks the inversion symmetry, and has different chiralities \cite{Greiter10, Quella18, Mila19}. The Affleck-Kennedy-Lieb-Tasaki (AKLT) state and its parent Hamiltonian have been constructed \cite{Greiter07a, Greiter07b, Katsura08}, and this system is found protected by the $Z_3\times Z_3$ symmetry with a doubly degenerate ground state \cite{Furusaki14}. We give an illustration of the AKLT structure under this representation in Fig. \ref{fig:aklt}, where the decomposition of the adjoint representation gives rise to two inequivalent virtual states, which leads to two inversion-breaking topological states.

On the experimental front, the cold atomic system has become a promising candidate for realizing various SPT phases in their SU(2) representations \cite{Xu, Katsura21}. For their SU($N$) counterparts, progress has been pushed forward to the alkali-earth atomic systems, as a natural emergence of SU($N$) symmetry by their nuclear spins \cite{Rey}. Since all the experimentally accessible spinful alkali-earth atoms are fermionic, proposals in this system are focusing on the fermionic SPT phases with local SU($N$) symmetry \cite{Totsuka13, Totsuka15, Momoi, Lecheminant19}, where most cases are with $N$ even and a unique ground state.

In this work, we take on an alternative route by considering a bosonic system, and show how to simulate the non-trivial Haldane phase in the SU(3) adjoint representation. We will first map the system to a Schwinger bosons representation with two species of bosons $a$ and $b$, then determine the ground-state phase diagram, and discuss its edge modes and physics at the dimer phase.

\section{The bosonic mapping}
We first map the SU(3) AFH chain with local adjoint representation to a bosonic model, starting from the Hamiltonian \cite{Auerbach}
\begin{align}
H_S=2\sum_{\langle ij\rangle,m} T_i^mT_{j}^m,
\end{align}
where $T_i^m$ is the $m$-th generator ($m$=1,2,\ldots,8) of the SU(3) adjoint representation at site $i$.

\begin{figure}[h]
    \centering
    \includegraphics[width=0.15\textwidth]{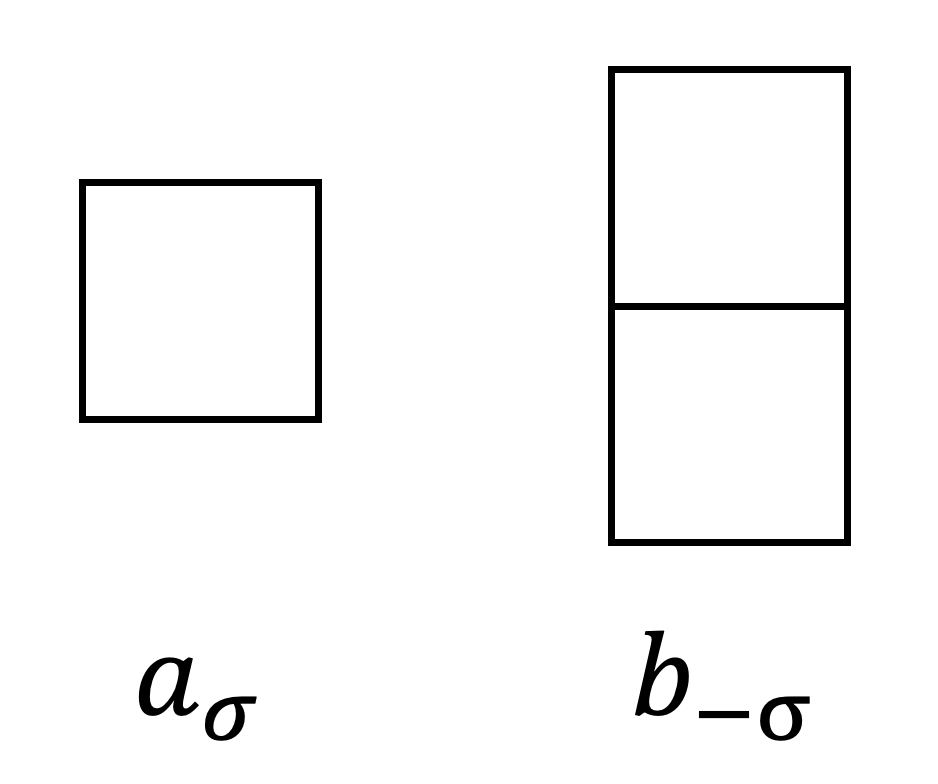}
    \caption{Two sets of Schwinger bosons that label the fundamental representation of SU(3) symmetry (shown as their corresponding Young diagrams). Three different internal hyperfine states ($\sigma=2, -2, 0$ as illustrated in Fig.~\ref{fig:8}b of each boson $a_\sigma$ and $b_\sigma$ are included to construct the triplet (quark) and antitriplet (antiquark) representations.}
    \label{fig:map}
\end{figure}

Unlike fermionic systems, there is no way to build antisymmetric representations from Schwinger bosons \cite{Greiter07b}. Instead, here we use two sets of Schwinger bosons to label the SU(3) fundamental representations as in Fig.~\ref{fig:map}. Then the generators of any SU(3) irreducible representation can be constructed by these two sets of bosons $a$ and $b$ (the quark and antiquark states) as \cite{Mathur01, Mukunda, Mathur09}
\begin{align}
T^m=\frac{1}{2}\sum_{\sigma\sigma^\prime}\left(a_\sigma^\dagger\lambda^m_{\sigma\sigma^\prime}a_{\sigma^\prime}+b_{-\sigma}^\dagger\bar{\lambda}^m_{\sigma\sigma^\prime}b_{-\sigma^\prime}\right),
\label{eq:boset}
\end{align}
with $\sigma=2,-2, 0$ as illustrated in Fig.~\ref{fig:map}, where $\lambda^m$ are the Gell-Mann matrices that account for the matrix representation of a single quark, and for the antiquark we have $\bar{\lambda}^m=-(\lambda^m)^*$. The Casimirs
\begin{align}
n_a=\sum_\sigma a_\sigma^\dagger a_{\sigma}, n_b=\sum_\sigma b_\sigma^\dagger b_{\sigma}
\end{align}
define the number of quarks and antiquarks in this representation Eq. (\ref{eq:boset}), and determines the dimension of this representation.

At first sight, for $n_a=n_b=1$ Eq. (\ref{eq:boset}) seems trivially equivalent to the direct product of two independent quark and antiquark representations, and thus the same as a two-leg SU(3) spin chain model with fundamental and antifundamental representations on each leg. We prove that different from this two-leg spin chain model, in fact the generator Eq. (\ref{eq:boset}) naturally projects out the singlet state, and leads to exactly the adjoint representation of SU(3) symmetry.

To see this, we write out the singlet state in terms of our bosonic space
\begin{align}
\vert S\rangle=\frac{1}{\sqrt{3}} \sum_\sigma a^\dagger_\sigma b^\dagger_{-\sigma}\vert vac\rangle.
\end{align}

The generator Eq. (\ref{eq:boset}) acting on the singlet state gives
\begin{align}
T^m \vert S\rangle &=\frac{1}{2\sqrt{3}}\sum_{\sigma\sigma^\prime}\left(a_\sigma^\dagger\lambda^m_{\sigma\sigma^\prime}b^\dagger_{-\sigma^\prime}+b_{-\sigma}^\dagger\bar{\lambda}^m_{\sigma\sigma^\prime}a^\dagger_{\sigma^\prime}\right) \vert vac\rangle\nonumber\\
&=\frac{1}{2\sqrt{3}}\sum_{\sigma\sigma^\prime}a_\sigma^\dagger\left(\lambda^m_{\sigma\sigma^\prime}+\bar{\lambda}^m_{\sigma^\prime\sigma}\right)b^\dagger_{-\sigma^\prime}\vert vac\rangle\nonumber\\
&=0,
\end{align}
which means that the generator $T^m$ projects out the singlet state, and there is no matrix element that couples the higher-dimensional representations to the singlet state. Thus for Casimirs $n_a=n_b=1$, our generator $T^m$ gives exactly the adjoint representation of SU(3) symmetry.

Under this representation, the SU(3) AFH spin chain becomes
\begin{align}
H_S=\sum_{\langle ij\rangle}H_{ij}^{aa}+H_{ij}^{ab}+H_{ij}^{ba}+H_{ij}^{bb},
\label{eq:mapham}
\end{align}
with
\begin{align}
H_{ij}^{aa}&=\frac{1}{2}\sum_{\sigma_i\sigma_i^\prime\sigma_j\sigma_j^\prime, m}a^\dagger_{i, \sigma_i}a^\dagger_{j, \sigma_j}a_{j, \sigma^\prime_j}a_{i, \sigma^\prime_i}\lambda^m_{\sigma_i \sigma^\prime_i}\lambda^m_{\sigma_j\sigma^\prime_j},\nonumber\\
 H_{ij}^{ab}&=\frac{1}{2}\sum_{\sigma_i\sigma_i^\prime\sigma_j\sigma_j^\prime, m}a^\dagger_{i, \sigma_i}b^\dagger_{j, \sigma_j}b_{j, \sigma^\prime_j}a_{i, \sigma^\prime_i}\lambda^m_{\sigma_i\sigma^\prime_i}\tilde{\lambda}^m_{\sigma_j\sigma^\prime_j},\nonumber\\
 H_{ij}^{ba}&=\frac{1}{2}\sum_{\sigma_i\sigma_i^\prime\sigma_j\sigma_j^\prime, m}b^\dagger_{i, \sigma_i}a^\dagger_{j, \sigma_j}a_{j, \sigma^\prime_j}b_{i, \sigma^\prime_i}\tilde{\lambda}^m_{\sigma_i\sigma^\prime_i}\lambda^m_{\sigma_j\sigma^\prime_j},\nonumber\\
 H_{ij}^{bb}&=\frac{1}{2}\sum_{\sigma_i\sigma_i^\prime\sigma_j\sigma_j^\prime, m}b^\dagger_{i, \sigma_i}b^\dagger_{j, \sigma_j}b_{j, \sigma^\prime_j}b_{i, \sigma^\prime_i}\tilde{\lambda}^m_{\sigma_i\sigma^\prime_i}\tilde{\lambda}^m_{\sigma_j\sigma^\prime_j}.	\nonumber
\end{align}

Carrying out the summation over different matrix indices and plugging into Eq.~(\ref{eq:mapham}), we have
\begin{align}
H_S=&\sum_{\langle ij\rangle,\sigma\sigma^\prime}\left(a_{i,\sigma}^\dagger a_{j,\sigma^\prime}^\dagger a_{j,\sigma}a_{i,\sigma^\prime}+b_{i,\sigma}^\dagger b_{j,\sigma^\prime}^\dagger b_{j,\sigma}b_{i,\sigma^\prime}\right.\nonumber\\
	&\left.-a_{i,\sigma}^\dagger b_{j,-\sigma}^\dagger b_{j,-\sigma^\prime}a_{i,\sigma^\prime}-b_{i,\sigma}^\dagger a_{j,-\sigma}^\dagger a_{j,-\sigma^\prime}b_{i,\sigma^\prime}\right).
\end{align}
Note that in deriving the above equation, we have taken the strong-coupling limit that each site has only a single $a$ and $b$ boson. Thus, we have mapped the SU(3) AFH chain with the adjoint representation to the spin-2 bosonic model.

\section{The phase diagram}\label{sec:phase}
To connect with the experiment (see the appendix for an experimental proposal), we consider the following Hamiltonian
\begin{align}
H&=\sum_{\langle ij\rangle,\sigma\sigma^\prime}\left[t\left(a^\dagger_{i,\sigma}a^\dagger_{j,\sigma^\prime}a_{j,\sigma}a_{i,\sigma^\prime}+b^\dagger_{i,\sigma}b^\dagger_{j,\sigma^\prime}b_{j,\sigma}b_{i,\sigma^\prime}\right)\right.\nonumber\\
&\left.-g_0\left(a^\dagger_{i,\sigma}b^\dagger_{j,-\sigma}b_{j,-\sigma^\prime}a_{i,\sigma^\prime}+b^\dagger_{i,-\sigma}a^\dagger_{j,\sigma}a_{j,\sigma^\prime}b_{i,-\sigma^\prime}\right)\right],
\label{eq:ham}
\end{align}
where the summation is over the nearest sites and $\sigma=0, \pm 2$ corresponds to different hyperfine states.
We note the spin-2 bosons are crucial, as the Clebsch-Gordan coefficients for two spin-1 atoms will lead to a sign change for different scattering states \cite{Ueda}, which will break the quark-antiquark singlet state. 
Here the positive intraspecies interaction $t$ accounts for the suppression of hopping for the quarks or antiquarks on the nearest sites, and the $g_0$ term comes from the scattering at total hyperfine spin $F=0$ channel, which keeps the quark-antiquark pair in the singlet space.

\begin{figure}[t]
    \centering
    \includegraphics[width=0.34\textwidth]{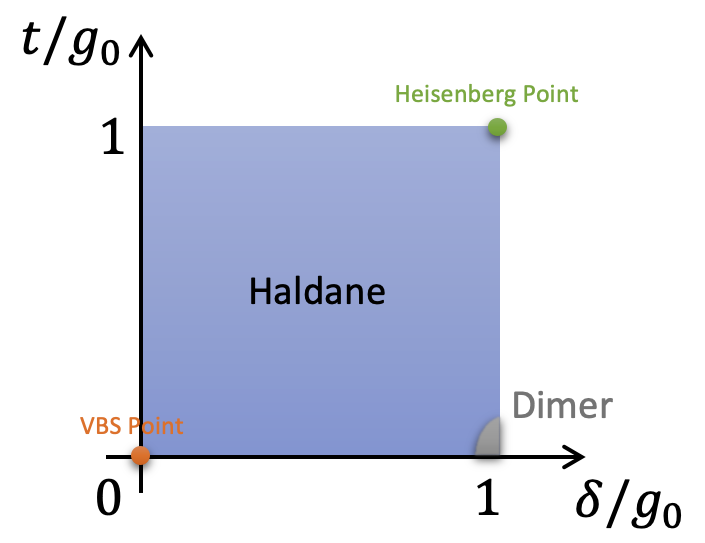}
    \caption{Characteristic phase diagram in our parameter regime. The Heisenberg limit and VBS limit in the Haldane phase are shown as the green and orange points. As explained in the Appendix, the experimentally accessible regime is for $t$ positive. The regime at $\delta/g_0>1$ is not shown, as it corresponds to its inverse chiral counterpart, and the phase diagram is symmetric along the line $\delta=g_0$.}
    \label{fig:phase}
\end{figure}

We show in Fig.~\ref{fig:phase} the characteristic phase diagram in the parameter regime at the strong-coupling limit. To illustrate the evolution to the SU(3) chiral Haldane phase, we include an energy shift $\delta$ to the left and right chiral states in the interspecies interaction as $-\sum_{\langle ij\rangle,\sigma\sigma^\prime}\left(g_0a^\dagger_{i,\sigma}b^\dagger_{j,-\sigma}b_{j,-\sigma^\prime}a_{i,\sigma^\prime}+\delta b^\dagger_{i,-\sigma}a^\dagger_{j,\sigma}a_{j,\sigma^\prime}b_{i,-\sigma^\prime}\right)$ with $j>i$ and $\delta/g_0 \in [0,1]$. This corresponds to a shift of these two species-dependent optical lattices in the experiment, and can be realized by including an additional trap along the $x$ axis. As expected, a well-defined valence bond solid (VBS) state (orange) is found at the point $t=\delta=0$ where a right chiral SU(3) singlet pair is formed at nearest sites. By increasing $t$ and $\delta$ the system can adiabatically evolve to the Heisenberg point (green) with $t=\delta=g_0$, where the Haldane phase shows double degeneracy of the left and right chiral states. As will be explained later, by decreasing $t$ from the Heisenberg point, the Haldane phase will become unstable, and there will be a phase transition to a dimer phase (gray).

To confirm the above qualitative phase diagram in Fig.~\ref{fig:phase}, we consider the following string order parameter in the Schwinger bosons representation
\begin{align}
C^{\rm Str}(i,j)=\left\langle \Delta n_{2}^ie^{i\pi\sum_{i<k<j}\Delta n_{2}^{k}}\Delta n_{2}^j\right\rangle,
\label{eq:str}
\end{align}
where $\Delta n_{2}^i=n_{a_2}^i-n_{b_{-2}}^i$ measures the particle number difference between the quark and antiquark state $a_2$ and $b_{-2}$ at site $i$. This number difference corresponds to the charge $Q^i=\Delta n_{2}^i=T^i_3+Y^i/2$ according to the Gell-Mann-Nishijima formula, and thus takes values $-1$, $0$, and $1$ in the meson state \cite{Greiner, Georgi}, where
\begin{align}
T^3&=\frac{1}{2}\left(n_{a_2}-n_{a_{-2}}\right)+\frac{1}{2}\left(n_{b_2}-n_{b_{-2}}\right),\\
Y&=n_{a_{2}}+n_{a_{-2}}-n_{b_{2}}-n_{b_{-2}}.
\end{align}
We note the string order we define here is similar to the one by Morimoto {\it et al.} \cite{Furusaki14}, and different from \cite{Quella12} where general Cartan operators are used.

\begin{figure}[ht]
    \centering
    \includegraphics[width=0.4\textwidth]{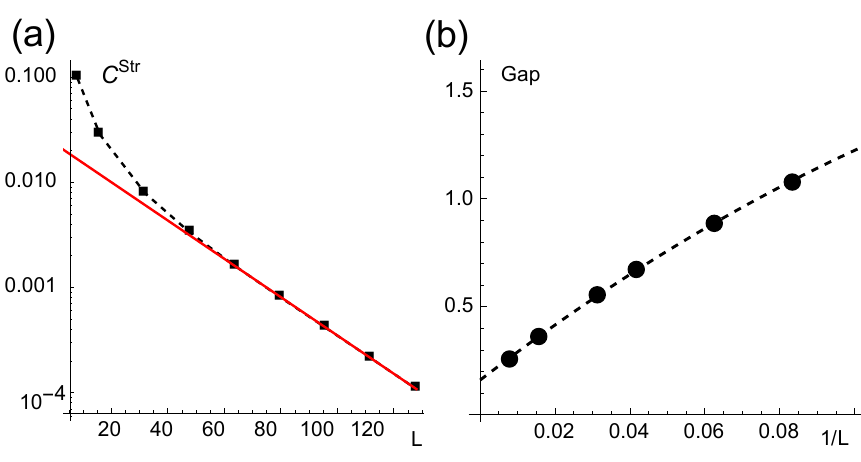}
    \caption{(a) The log-linear plot of the ground state string order at the dimer point, which shows an exponential behavior at long distances as indicated by the red line, with the fitting giving $C^{Str}(r)=0.023\times e^{-0.08r}$. (b) The energy gap at the dimer point as a function of inverse system size $1/L$. The dashed line corresponds to a second-order polynomial fitting, which gives an energy gap of around 0.16.}
    \label{fig:gap}
\end{figure}

The existence of a new phase besides the Haldane one can be revealed by the string order at the dimer point as shown in Fig.~\ref{fig:gap}a, where we give a log-linear plot of $C^{Str}$ as a function of system size. We consider system sizes as large as $L=128$ with the bond dimension as large as $m=2000$. For long distances, the string order shows an exponential behavior as guided by the red line, which is consistent with a trivial gapped phase. We then further calculate the energy gap for various system sizes as shown in Fig.~\ref{fig:gap}b. We do a second-order polynomial fitting to the energy gap as a function of $1/L$ as the dashed line, and determine the energy gap at the dimer point to be around 0.16.

To give the characteristic phase diagram as shown in Fig.~\ref{fig:phase}, we carry out calculations about the string order defined in Eq. (\ref{eq:str}) with an additional dimer order
\begin{align}
C^{dim}=\langle \Delta n_{2}^{L/2}\Delta n_{2}^{L/2+1}-\Delta n_{2}^{L/2}\Delta n_{2}^{L/2-1}\rangle,
\end{align}
which measures the difference of correlations at even and odd bonds. The dimer order $C^{dim}$ can measure the translational symmetry-breaking state that appears in the system.

\begin{figure}[t]
    \centering
    \includegraphics[width=0.35\textwidth]{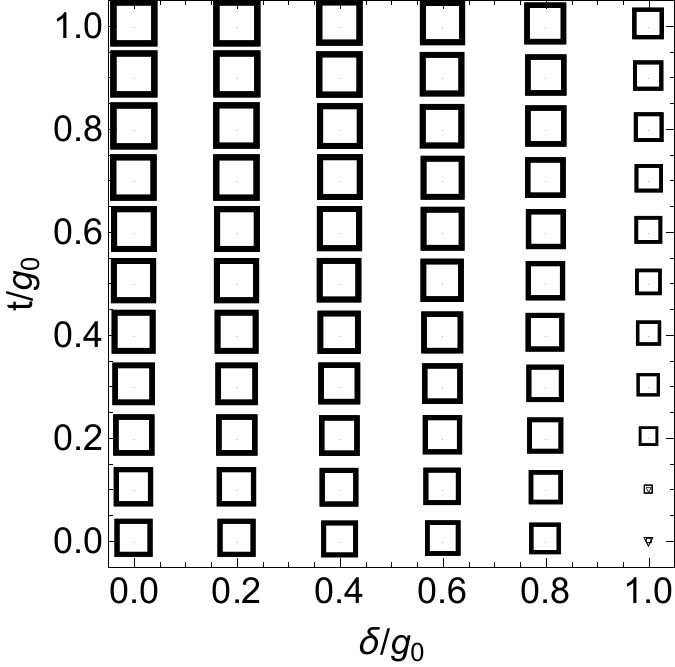}
    \caption{The order parameters obtained from DMRG calculations with system size $L=128$. The squares and triangles label the string and dimer orders respectively, with their sizes corresponding to their magnitudes. A translational symmetry-breaking dimer phase is observed at a small regime near the dimer point $t/g_0=0, \delta/g_0=1.0$.}
    \label{fig:string}
\end{figure}

For these two oder parameters, we carry out DMRG calculations with system size $L=128$ and bond dimension as large as 1600. Our results are summarized in Fig.~\ref{fig:string}, where the sizes of squares and triangles label the magnitudes of the string and dimer orders respectively. We find there is a rather small regime near the dimer point $t/g_0=0, \delta/g_0=1.0$ where the system breaks the translational symmetry and goes into a dimer phase. We note a similar phase transition has also been observed in a two-leg SU(3) spin ladder, where a similar case at $\delta/g_0=1$ in our Fig.~\ref{fig:string} has been considered \cite{Totsuka20}. These results confirm our qualitative phase diagram in Fig.~\ref{fig:phase}.

\section{Identifying the edge modes}\label{sec:observation}
The most direct experimental signature of the Haldane phase would be observing its edge modes. In a lattice system, the quantum gas microscope is a powerful technique that can give direct local observations, as well as correlations \cite{Ott, Bakr09, Bakr10, Kuhr10, Kuhr11, Sompet}. To connect with these actual experiment, we focus on a limited lattice size with $L=16$.
 
\begin{figure*}[h]
    \centering
    \includegraphics[width=.88\textwidth]{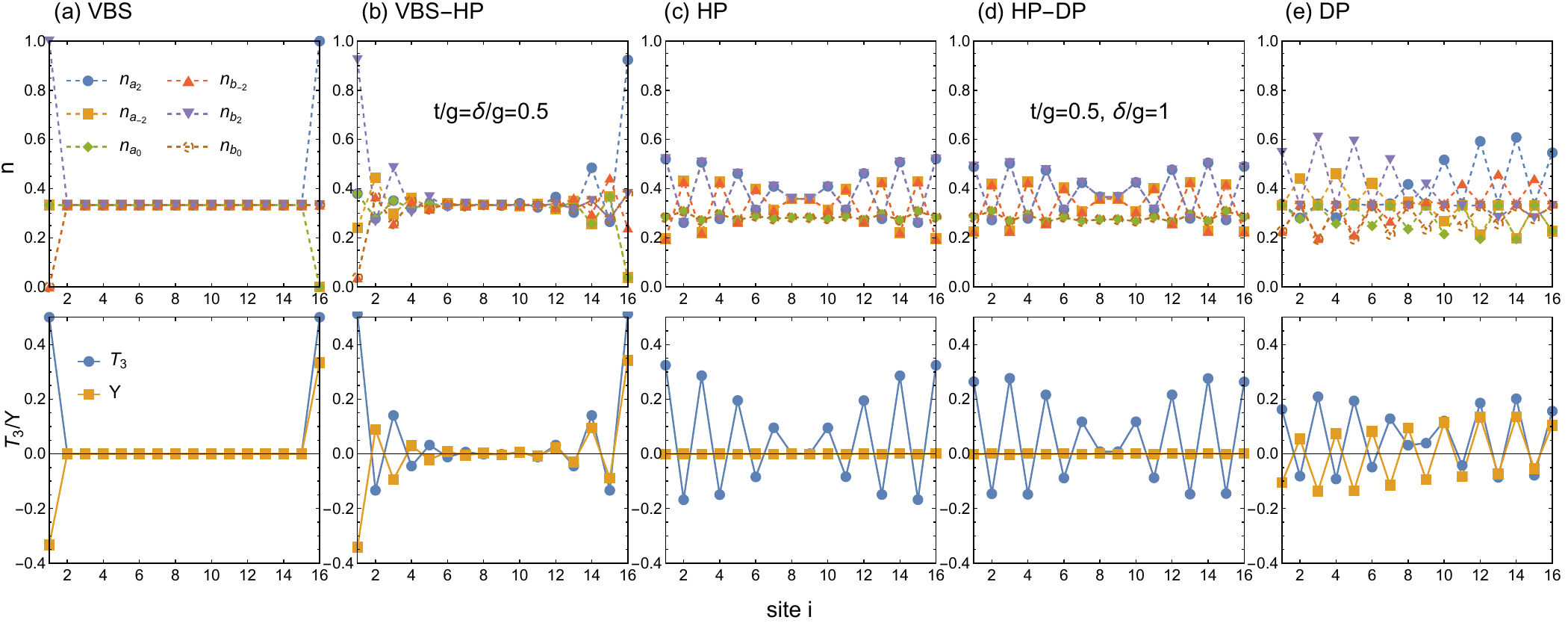}
    \caption{The local particle number distribution and the corresponding local $T_3$ and $Y$ for different parameters. From (a) to (e) we are at the VBS point ($t/g=\delta/g=0$), VBS to Heisenberg point (VBS-HP) ($t/g=\delta/g=0.5$), Heisenberg point (HP) ($t/g=\delta/g=1$), Heisenberg to dimer point (HP-DP) ($t/g=0.5, \delta/g=1$), and the dimer point (DP) ($t/g=0, \delta/g=1$). The edge modes can be clearly seen in both VBS and VBS-HP, which supports a chiral Haldane phase. In the HP, the Haldane phase shows two-fold degeneracy, and as a result the particle distribution shows edge modes of two different chiral states, and in the same sense, the local $Y$ cancels to zero due to the superposition of these two chiral Haldane phases. The HP-DP in (d) is close to the transition, as the edge modes become weaker. However, since the local $Y$ is still homogeneous and the double degeneracy is conserved, this state is closer to the Haldane phase. In (e) we observe a spontaneous breaking of lattice translational symmetry from the local $Y$, and thus this state belongs to a trivial dimer phase, with the double degeneracy in particle distribution also broken.}
    \label{fig:dens}
\end{figure*}

We consider an open chain lattice, and to reduce the Hilbert space and lift up additional degeneracies, we restrict the subspaces to have $T_3=1$ and $Y=0$. This corresponds to post-select the microscopic images with Hilbert space $S_z=2$ for both $a$ and $b$ particles.

We show the particle number distribution and its corresponding local $T_3$ and $Y$ for different parameters in Fig.~\ref{fig:dens}. The VBS state in Fig.~\ref{fig:dens}a clearly shows a single particle $a_2$ and $b_2$ appear at each edge of the open chain, which signals the quark and antiquark edge modes that appear in the chiral Haldane phase. These edge modes can also be seen from the $T_3/Y$ plot. In Fig.~\ref{fig:dens}b we show the position between VBS and Heisenberg point (HP) at $t/g=\delta/g=0.5$. We observe the same edge modes that are robust in the Haldane phase.

For the HP in Fig.~\ref{fig:dens}c, we observe different behaviors. The edge modes become weak and we observe two edge $a_2$ and $b_2$ modes on both sides. This signals a non-trivial property of the Haldane phase in the SU(3) adjoint representation, that the ground state can support a two-fold degeneracy. The superposition of these two degenerate ground states with opposite chiralities leads to the cancellation of local $Y$ as illustrated in Fig.~\ref{fig:dens}c, while the edge modes are more visible from the local $T_3$ plot. The decay of edge modes as a function of sites becomes slower, indicating a smaller energy gap at the HP.

At the HP to dimer point (DP) in Fig.~\ref{fig:dens}d for $t/g=0.5$ and $\delta/g=1$, these edge modes are even weaker, and the system is close to a transition to the dimer phase. However, the $Y$ plot still shows a translational symmetry, indicating that the system is closer to a Haldane phase. At the DP in Fig.~\ref{fig:dens}e, we observe a spontaneous breaking of lattice translational symmetry from the $Y$ plot, and the system is in a trivial dimer phase, where the doubly degenerate edge modes are also lost from the local particle density.

\section{Physics at the dimer phase}\label{sec:dimer}
\begin{figure}[t]
    \centering
    \includegraphics[width=0.4\textwidth]{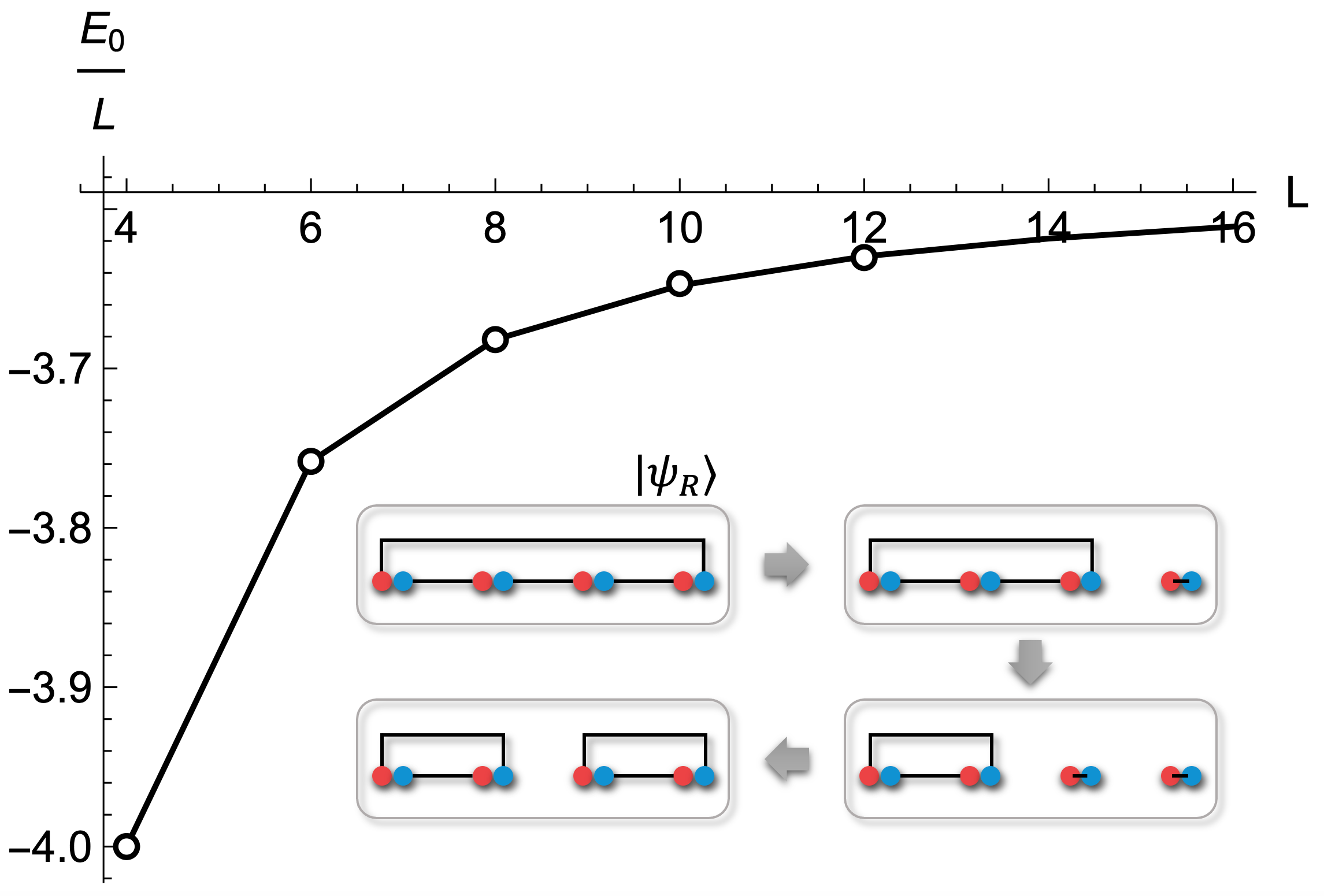}
    \caption{Ground state energy density $E_0/L$ at the dimer point for different system sizes $L$. The open points are results from the coupled Eqs. (\ref{eq:coupledeq}), and the solid line shows the DMRG results. The inset illustrates four substates in the ground state of the $L=4$ system, with blue and red points labeling the quark and antiquark states and solid lines representing the SU(3) singlet bonds.}
    \label{fig:dimer}
\end{figure}

The physics of this dimer phase is not trivial and lacks analytical solutions. To further understand this phase, we consider the following reduced Hamiltonian
\begin{equation*}
	H_d=-\sum_{\langle ij\rangle,\sigma\sigma^\prime}\left(a^\dagger_{i,\sigma}b^\dagger_{j,-\sigma}b_{j,-\sigma^\prime}a_{i,\sigma^\prime}+b^\dagger_{i,-\sigma}a^\dagger_{j,\sigma}a_{j,\sigma^\prime}b_{i,-\sigma^\prime}\right)
\end{equation*}
and try to construct an analytical ground state ansatz, starting from the right chiral Haldane phase $\vert\psi_R\rangle$ as illustrated in the inset of Fig.~\ref{fig:dimer}, where the red and blue dots label the quark and antiquark states at site $i$, and the solid lines label the SU(3) singlet bonds. We observe that if the Hamiltonian $H_d$ is acting on $\vert\psi_R\rangle$, the resulting states are a superposition of states with each quark/antiquark state connected with a single solid line. So the effects of the Hamiltonian $H_d$ just shift these solid lines. This observation suggests the ground state of $H_d$ might be constructed from states with fully connected singlet states.

We label each substate as $\vert\{r_i\}\rangle$ with the quark state at site $i$ connected to the antiquark state at site $r_i$, and different structures correspond to different permutations of the list from $1$ to $L$. For example, the right chiral Haldane phase $\vert\psi_R\rangle$ can be labeled as $\vert 2,3,4,\cdots, L,1\rangle$. We take into account the translational invariance of the system and consider the periodic boundary condition, and it is more convenient to transform the Hamiltonian into momentum space and the ground state ansatz can be written as
\begin{align}
\vert\psi\rangle=&\sum_{\{r_i\}}c_{\{r_i\}}\vert r_{1}r_{2}\cdots \rangle\nonumber\\
=&\sum_{\{r_i\}}c_{\{r_i\}}\cdot\sum_{k_1,\sigma_1,k^\prime_1,\sigma^\prime_1,\cdots }a^\dagger_{k_1,\sigma_1}b^\dagger_{k^\prime_1,\sigma^\prime_1}a^\dagger_{k_2,\sigma_2}b^\dagger_{k^\prime_2,\sigma^\prime_2}\cdots\nonumber\\
&\times e^{i\left(k_1+k^\prime_1r_{1}\right)}e^{i\left(2k_2+k^\prime_2r_{2}\right)}\cdots \vert vac\rangle,
\end{align}
where the summation is over all possible configurations of the SU(3) singlet bonds labeled as $\{r_i\}$. 
The coefficients $c_{\{r_i\}}$ can be determined through
\begin{align}
&H_d\vert r_{1}r_{2}\cdots \rangle\nonumber\\
=&-N\sum_i\left(\delta_{i-r_{i}-1,0/-L}+\delta_{i-r_{i}+1,0/L}\right)\vert \cdots r_{i}\cdots \rangle\nonumber\\
&-\sum_{i\neq j}\left(\delta_{i-r_{j}-1,0/-L}+\delta_{i-r_{j}+1,0/L}\right)\vert \cdots r_{j}\cdots r_{i}\cdots\rangle\nonumber\\
=&-N\sum_i\left(\delta_{i-r_{i}-1,0/-L}+\delta_{i-r_{i}+1,0/L}\right)\vert r_{1}r_{2}\cdots \rangle\nonumber\\
&-\sum_{i\neq j}\left(\delta_{i-r_{j}-1,0/-L}+\delta_{i-r_{j}+1,0/L}\right)\vert \cdots r_{j}\cdots r_{i}\cdots\rangle,\nonumber
\end{align}
where $N=3$ for a SU(3) system. Thus we get the coupled equations for the coefficients
\begin{align}
&Ec_{r_1r_2\cdots}\nonumber\\
=&-N\sum_i\left(\delta_{i-r_{i}-1,0/-L}+\delta_{i-r_{i}+1,0/L}\right)c_{r_1r_2\cdots }\nonumber\\
&-\sum_{i\neq j}\left(\delta_{i-r_{i}-1,0/-L}+\delta_{i-r_{i}+1,0/L}\right)c_{\cdots r_j\cdots r_i\cdots }\nonumber\\
=&-\sum_{ij}\left[1+\left(N-1\right)\delta_{ij}\right]\nonumber\\
&\times\left(\delta_{i-r_i-1,0/-L}+\delta_{i-r_i+1,0/L}\right)c_{\cdots r_j\cdots r_i\cdots }.
\label{eq:coupledeq}
\end{align}

The above coupled equations are huge and hard to solve. For system size $L$, there are at most $L!-1$ coefficients to find. However, we can always simplify these equations by the symmetry of the subspaces. For example, we always have the coefficient of right chiral Haldane phase $\vert 2,3,4,\cdots, L,1\rangle$ equals the one of the left chiral Haldane phase $\vert L,1,2,.., L-1\rangle$. The coupled equations can then be solved as follows. We start from the left and right chiral Haldane phases with other fully connected states unoccupied. After finding the energy $E$, we take into account the equations with states coupled with the occupied ones and find the corresponding coefficients. The energy is then updated with new occupied states. This procedure is carried out iteratively until we get the convergent ground-state energy. In Fig.~\ref{fig:dimer} we show the ground state energy densities with above ansatz as the open points, which coincide well with our DMRG results (solid line). This justifies our description at the dimer point.

We can now understand why the positive $t$ in Eq. (\ref{eq:ham}) is crucial to realize the SU(3) Haldane phase, since this term suppresses the hopping of quark/antiquark states that leads to the product states as shown by the gray arrows in Fig.~\ref{fig:dimer}, and thus stabilizes the Haldane phase.

\section{Summary and outlook}
To conclude, we have provided a Schwinger boson realization of the non-trivial Haldane phase with local SU(3) adjoint representation. We identify a phase transition from the Haldane phase to a dimer phase, and show evidence of the topological edge modes. We also explain the physics at the dimer phase by constructing the ground state wavefunction at the dimer point.

There are also experimental challenges. Due to the existence of an extremely small energy gap in the SU(3) Haldane phase \cite{Mila20, Verstraete}, one might expect the ground state to be very sensitive to the temperature. However, as the experimentally accessible regime could be closer to the VBS point, the excited states will be lifted compared to the Heisenberg point, resulting in a larger energy gap. Another possible way to overcome this is to consider a finite-temperature topology that can detect the non-trivial ground state by means of its thermal ensemble, for example, to measure its ensemble geometrical phase \cite{Diehl, Cooper}.

\section{Appendix: The experimental proposal}
\label{app:exp}
\begin{figure}[h]
    \centering
    \includegraphics[width=0.4\textwidth]{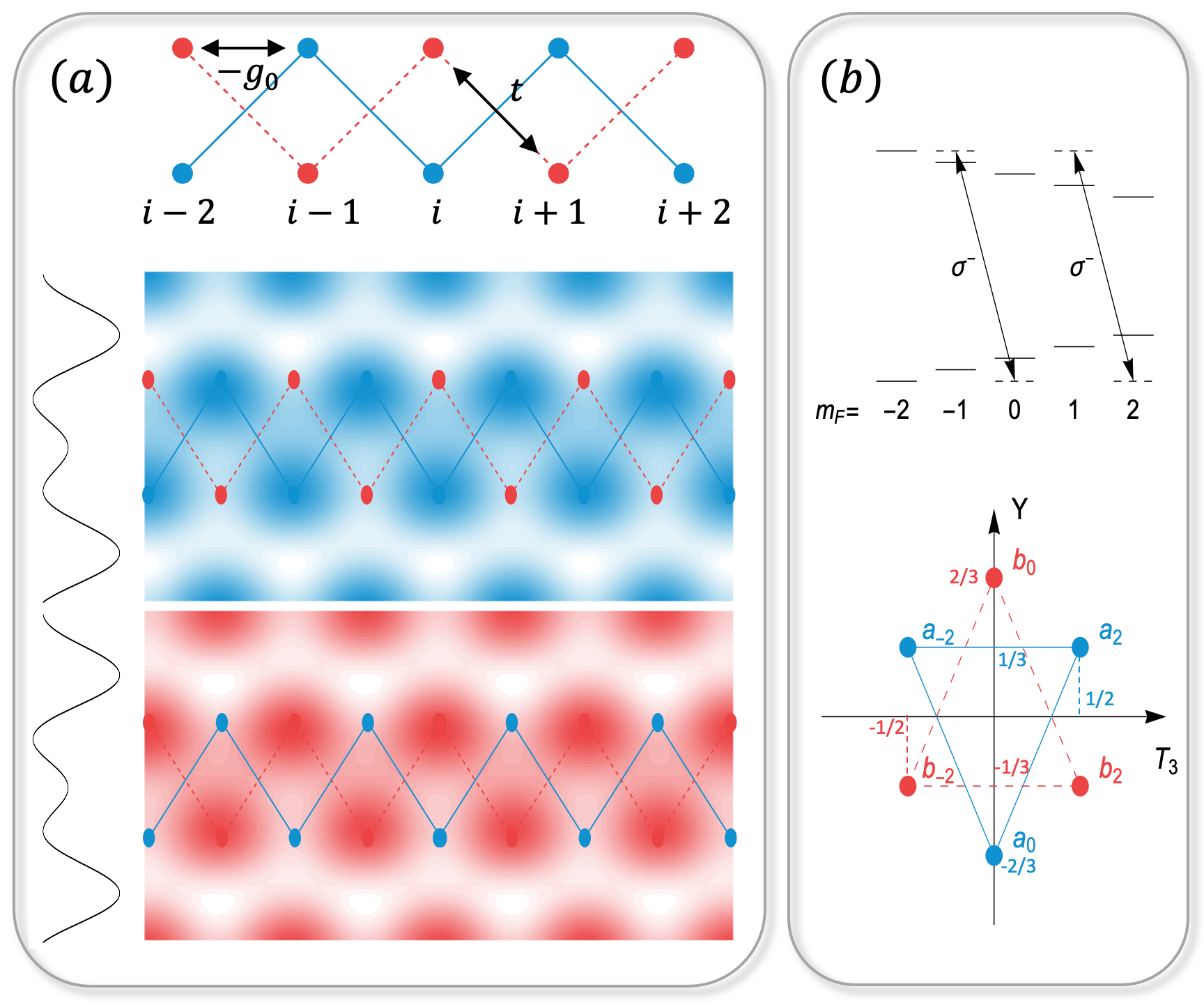}
    \caption{An illustration of the experimental implementation. (a) The species-dependent optical potential is used to construct two interlaced zigzag lattices, where each species is composed of three degenerate states in the hyperfine $F=2$ state. (b) By combining the magnetic field and $\sigma^-$ polarized lasers, the three lowest hyperfine states in the spin-2 bosons $a$ and $b$ are picked up, realizing the quark and antiquark representations of the SU(3) symmetry.}
    \label{fig:8}
\end{figure}

The experimental setup can be illustrated in Fig.~\ref{fig:8}. We create a species-dependent zigzag optical lattice by combining the use of lasers with three different wavelengths: $\lambda_0$ which is blue detuned from species $a$ and red detuned from species $b$, $ \lambda_r=\sqrt{2}\lambda_0$ which is red detuned from the two species, and $\lambda_b=\lambda_0/\sqrt{2}$ which is blue detuned from the two species.
Two sets of counter propagating $\lambda_0$ lasers create a species dependent diagonal square optical lattice $V_{dep}=V_0\cos(\frac{k_0x+k_0y}{\sqrt{2}})^2+V_0\cos(\frac{k_0x-k_0y}{\sqrt{2}})^2$ with $k_0=2\pi/\lambda_0$. A periodic modulation is added along the $y$ axis to separate each two-leg system and reduce the interspecies spin-mixing interaction at the same site with $V_{y}=V_0\cos(k_by+\pi/2+\phi)^2-V_0\cos(k_ry+\pi/4)^2$, as illustrated in the solid curve in Fig.~\ref{fig:8}a, where we have added a small phase difference $\phi$ to create a tilted lattice. A demonstration of the realized species-dependent optical lattice for atoms $a$ (blue) and $b$ (red) is shown in Fig.~\ref{fig:8}a, where the denser color corresponds to a lower trap depth.
Here the tilted lattices prevent the hopping of atoms at the nearest sites within each species.
To realize a SU(3) symmetric system within the hyperfine spin $F=2$ states, we propose to combine the magnetic field with $\sigma^-$ polarized lasers to pick up the lowest hyperfine states with $m_F=0,\pm2$ as shown in Fig.~\ref{fig:8}b. In this case, we can map these lowest hyperfine states $a_\sigma$ and $b_\sigma$ to the quark and antiquark states.

\acknowledgements
This research is supported by the Fundamental Research Funds for the Central Universities (No. FRF-BD-25-033). We would like to thank Philippe Lecheminant, Thomas Quella, and Hosho Katsura for helpful comments.

\clearpage
\onecolumn

\begin{center}
\textbf{\large Supplementary Material: Spinor bosons realization of the SU(3) Haldane phase with adjoint representation}
\end{center}

\newcounter{suppref}
\renewcommand{\thesuppref}{[S\arabic{suppref}]}
\newcommand{\citeSupp}[1]{\ref{#1}}

\setcounter{equation}{0}
\setcounter{figure}{0}
\setcounter{page}{1}
\renewcommand{\thepage}{S\arabic{page}}     
\renewcommand{\thefigure}{S\arabic{figure}}
\renewcommand{\theequation}{S\arabic{equation}}

\section{Realize a positive intraspecies interaction $t$}
\begin{figure}[ht]
    \centering
    \includegraphics[width=0.4\textwidth]{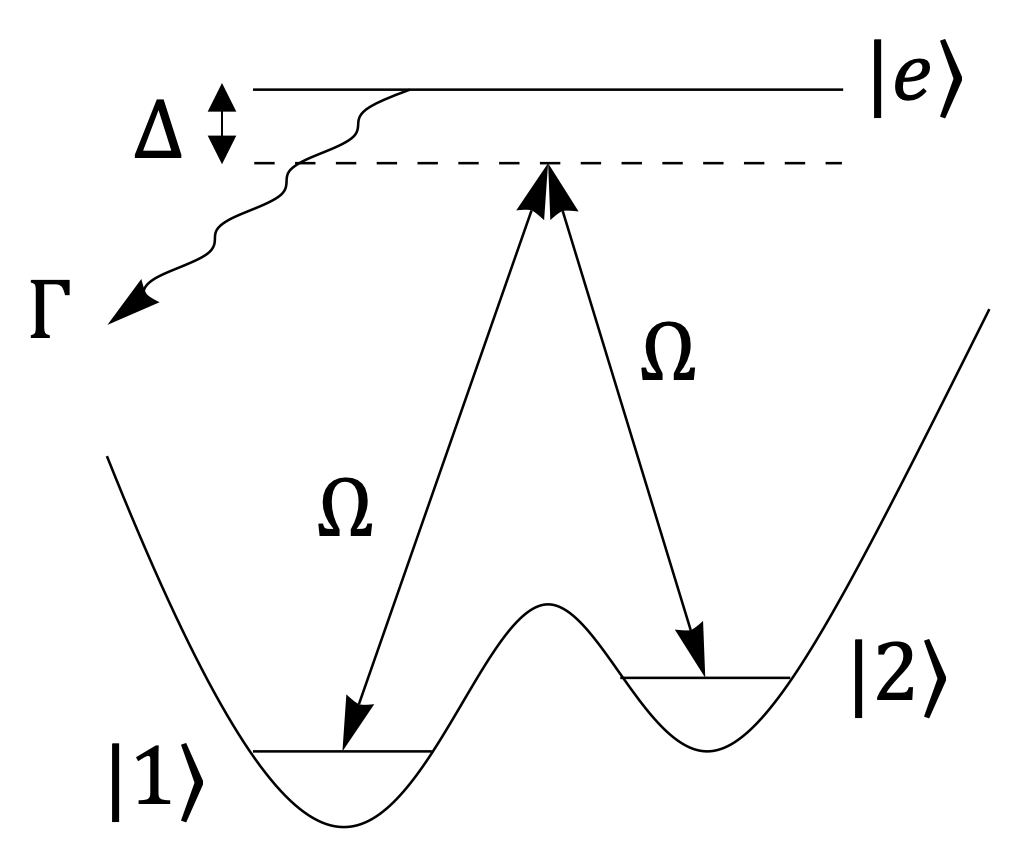}
    \caption{By coupling the same hyperfine state at the tilted lattice to a higher excited state $\vert e\rangle$ with a large detuning $\Delta$ and decay rate $\Gamma$, we end up with an effective imaginary hopping.}
    \label{fig:S1}
\end{figure}

We propose to use a two-photon Raman process to realize an effective imaginary hopping between tilted lattices as shown in Fig.~\ref{fig:S1}.

For simplicity, let's consider a single particle with the same hyperfine state $m_F=0, 2$ or $-2$ at the nearest-neighbor lattice site labeled as $\vert 1\rangle$ and $\vert 2\rangle$, and the excited state $\vert e\rangle$ has a decay rate $\Gamma$ and detuning $\Delta$. The Hamiltonian of this two-site system can be written as
\begin{align}
H_{12}=\Omega(b_1^\dagger b_e+b_2^\dagger b_e+{\rm H.c.})+(\Delta-i\Gamma) b_e^\dagger b_e,
\label{eq:h12}
\end{align}
where $\Omega$ labels the Rabi frequency of these two Raman lasers. If the excited state has a large detuning and decay rate ($\Delta, \Gamma\gg\Omega$), atomic numbers in this state will quickly decay into a negligible steady value. In this case, we can adiabatically eliminate the excited state by considering the equation of motion
\begin{align}
\dot{b}_e^\dagger=i[H_{12}, b_e^\dagger]=0,
\end{align}
which gives
\begin{align}
b_e^\dagger=-\frac{\Omega}{\Delta-i\Gamma}(b_1^\dagger+b_2^\dagger).
\end{align}

\begin{figure}[ht]
    \centering
    \includegraphics[width=0.5\textwidth]{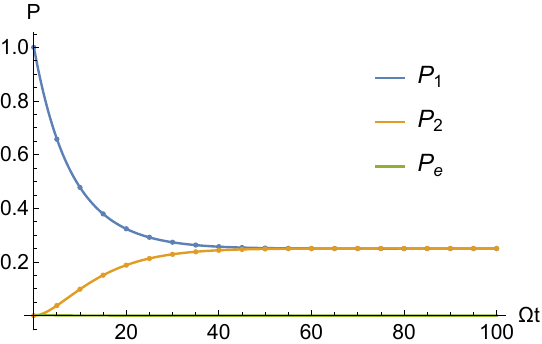}
    \caption{Probabilities of different states as a function of time starting from state $\vert 1\rangle$. The solid lines show the evolution of Hamiltonian $H_{12}$, and the points correspond to results from the effective Hamiltonian $\tilde{H}_{12}$. The detuning and decay rate are chosen as $\Delta/\Omega=5$ and $\Gamma/\Omega=20$.}
    \label{fig:S2}
\end{figure}

By eliminating the excited state $\vert e\rangle$ using the above equation, we get an effective Hamiltonian
\begin{align}
\tilde{H}_{12}=-\frac{\Omega^2}{\Delta-i\Gamma}(n_1+n_2+b_1^\dagger b_2+b_2^\dagger b_1).
\label{eq:effh12}
\end{align}

We compare this effective Hamiltonian with the original one in Fig.~\ref{fig:S2}, and illustrate the time evolution of different states starting from state $\vert 1\rangle$ at detuning $\Delta/\Omega=5$ and decay rate $\Gamma/\Omega=20$. From the dotted points we can find that the above effective Hamiltonian $\tilde{H}_{12}$ gives us a desirable description of the long-time steady state. The appearance of such a steady state is due to the fact that the non-Hermitian system supports a real eigenenergy solution. In this case, a Rayleigh-Schr\"odinger type non-Hermitian perturbation theory \citeSupp{Cohen} can be carried out.

Now considering a large decay rate $\Gamma\gg\Delta$, the Hamiltonian $\tilde{H}_{12}$ gives an imaginary hopping $i\gamma=-i\Omega^2/\Gamma$. Then under the non-Hermitian perturbation theory, and taking into account the strong-coupling limit with unit filling that prohibits double occupation at each lattice site, only the double hopping terms support a non-trivial contribution and we arrive at the following effective term
\begin{align}
H_t=t\sum_{\langle ij\rangle,\sigma\sigma^\prime}\left(a^\dagger_{i,\sigma}a^\dagger_{j,\sigma^\prime}a_{j,\sigma}a_{i,\sigma^\prime}+b^\dagger_{i,\sigma}b^\dagger_{j,\sigma^\prime}b_{j,\sigma}b_{i,\sigma^\prime}\right),
\end{align}
with the positive coefficient $t=\Omega^4/(\Gamma^2U)$ and $U$ the enegy shift for a single hopping.
S3

\section{String order}
\begin{figure}[ht]
    \centering
    \includegraphics[width=0.7\textwidth]{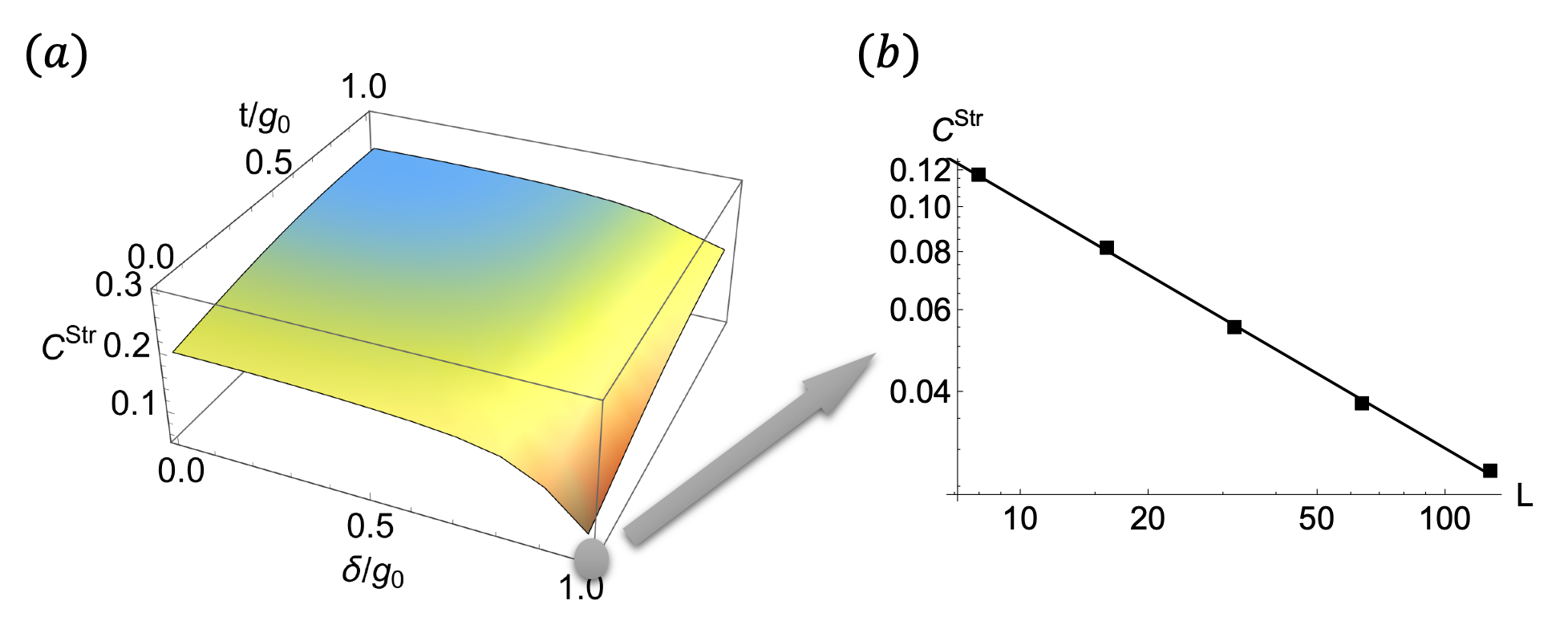}
    \caption{(a) The string order parameter $C^{\rm Str}=C^{\rm Str}(L/4,3L/4)$ as a function of intraspecies interaction $t$ and interspecies spin-mixing energy shift $\delta$. (b) Finite-size scaling of the string order parameter shows a power-law behavior at the dimer point for $t/g_0=0$ and $\delta/g_0=1$.}
    \label{fig:S3}
\end{figure}

The experimental observation of string order is within the current cold atoms techniques. However, the system sizes should be very limited. We show this string order parameter for different intraspecies interaction $t$ and interspecies spin-mixing energy shift $\delta$ for an experimentally realizable system size $L=16$ in Fig.~\ref{fig:S3}a by DMRG calculations. We consider an open system, and choose subspaces with $T_3=1$ and $Y=0$. We observe a finite string order all through our parameter regime, and find it decreases apparently close to $\delta/g_0=1$ and $t=0$, which is consistent with the existence of a quantum phase transition near this point at the thermodynamic limit. 

We note that due to the small system size in the experiment, it is unlikely to observe the transition to a dimer phase from the string order. We further push the DMRG calculations all through to system size $L=128$ and keep up to 1600 states at the dimer point, the string order parameter shows a power-law behavior as in Fig.~\ref{fig:S3}b, in contrast to the exponential behavior as expected for a gapped dimer phase. We attribute this to our selection of subspaces that rules out the dimer ground state.

\section{Entanglement spectrum}
\begin{figure}[t]
    \centering
    \includegraphics[width=0.7\textwidth]{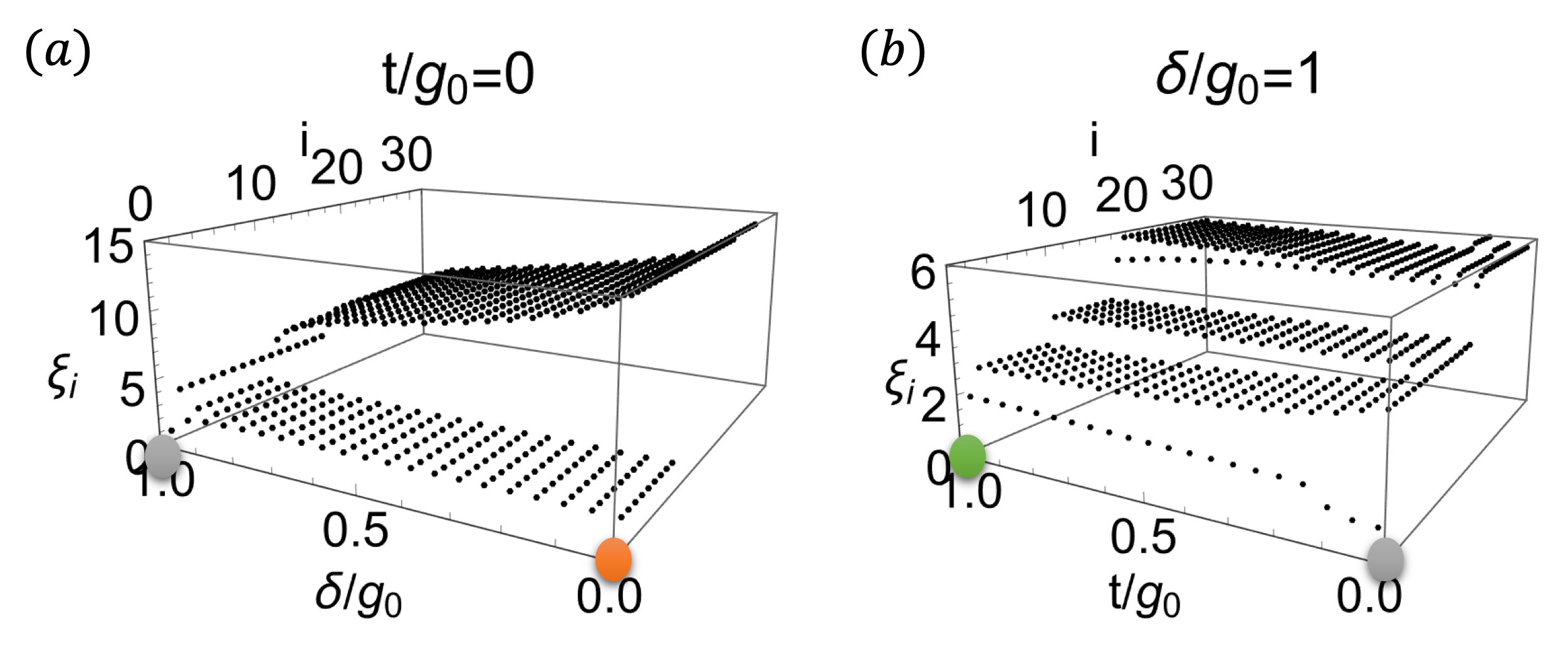}
    \caption{Evolution of the entanglement spectrum from the VBS to dimer point at $t/g_0=0$ in (a) and from the dimer to Heisenberg point at $\delta/g_0=1$ in (b). There is a spectrum gap above the 9th entanglement eigenvalue $\xi_9$ in both the VBS (orange) and Heisenberg (green) points. Approaching the dimer phase (gray) this spectrum gap closes.}
    \label{fig:S4}
\end{figure}

The experimental observation of the entanglement spectrum might still be challenging, however it provides a more robust test to symmetry-protected states \citeSupp{Pollmann}. We illustrate the evolution of the entanglement spectrum in Fig.~\ref{fig:S4} from the VBS and Heisenberg points to the dimer point. To avoid boundary effects, here we consider a periodic chain and take an experimentally accessible chain length $L=16$.

The entanglement spectrum shows a well-defined 9-fold degeneracy in the VBS point (orange) in Fig.~\ref{fig:S4}a, corresponding to the entanglement of two fundamental SU(3) edge modes (quark or antiquark states) with each of which appears at one boundary of the half system. This degeneracy clearly reveals the appearance of the chiral Haldane phase.
Besides this, we also observe a very large entanglement gap above the ninth entanglement eigenvalue in the Haldane phase. Close to the dimer phase (gray) around $t/g_0=0$ and $\delta/g_0=1$, the 9-fold degeneracy is lifted up with this entanglement gap vanishing.
The closing of this entanglement gap has an intuitive explanation. As the system goes into a trivial state, the edge modes can only support representations with integer numbers of Young diagram boxes, so the lowest entanglement eigenvalue would be one singlet at each boundary and unique, and the second lowest entanglement eigenvalues would be a singlet and an octet at each boundary with a total 16-fold degeneracy. This gives a closing of the entanglement gap above the ninth entanglement eigenvalue and signals the appearance of a trivial state.

As illustrated in Fig.~\ref{fig:S4}b, deviated from the dimer phase the entanglement gap above the ninth entanglement eigenvalue reopens, and the nine smallest entanglement eigenvalues gather when approaching the Heisenberg point (green).
Here we note our DMRG calculations naturally search for one of the two degenerate chiral ground states in the Heisenberg point, with the lowest entanglement spectrum showing 9-fold (instead of 18-fold) degeneracy approaching the thermodynamic limit.
In contrast to the VBS point, the Heisenberg point suffers much from the finite-size effect and shows an absence of the 9-fold degeneracy in its lowest entanglement eigenvalues. This finite-size effect is similar to the spin-1 Haldane phase, and is due to the spectrum gap between the SU(3) singlet state and 8-fold octet states formed by the two SU(3) fundamental representations appearing at the edges. We also find that the entanglement gap above the ninth entanglement eigenvalue is more robust than the 9-fold degeneracy of the lowest entanglement eigenvalues, and suffers little from the finite-size effect.

\section{The free bond excitations approximation}
In the main text, we give exact calculations of the dimer point and explain how the ground state supports the entanglement behavior. However, these exact calculations can only be limited to rather small system sizes. Here we provide a free bond excitations approximation that neglects the interaction between different excitations. This approximation, though not exact, can be pushed to very large system sizes, and provide an upper limit for the ground state energy of the dimer point.

\begin{figure}[ht]
    \centering
    \includegraphics[width=0.7\textwidth]{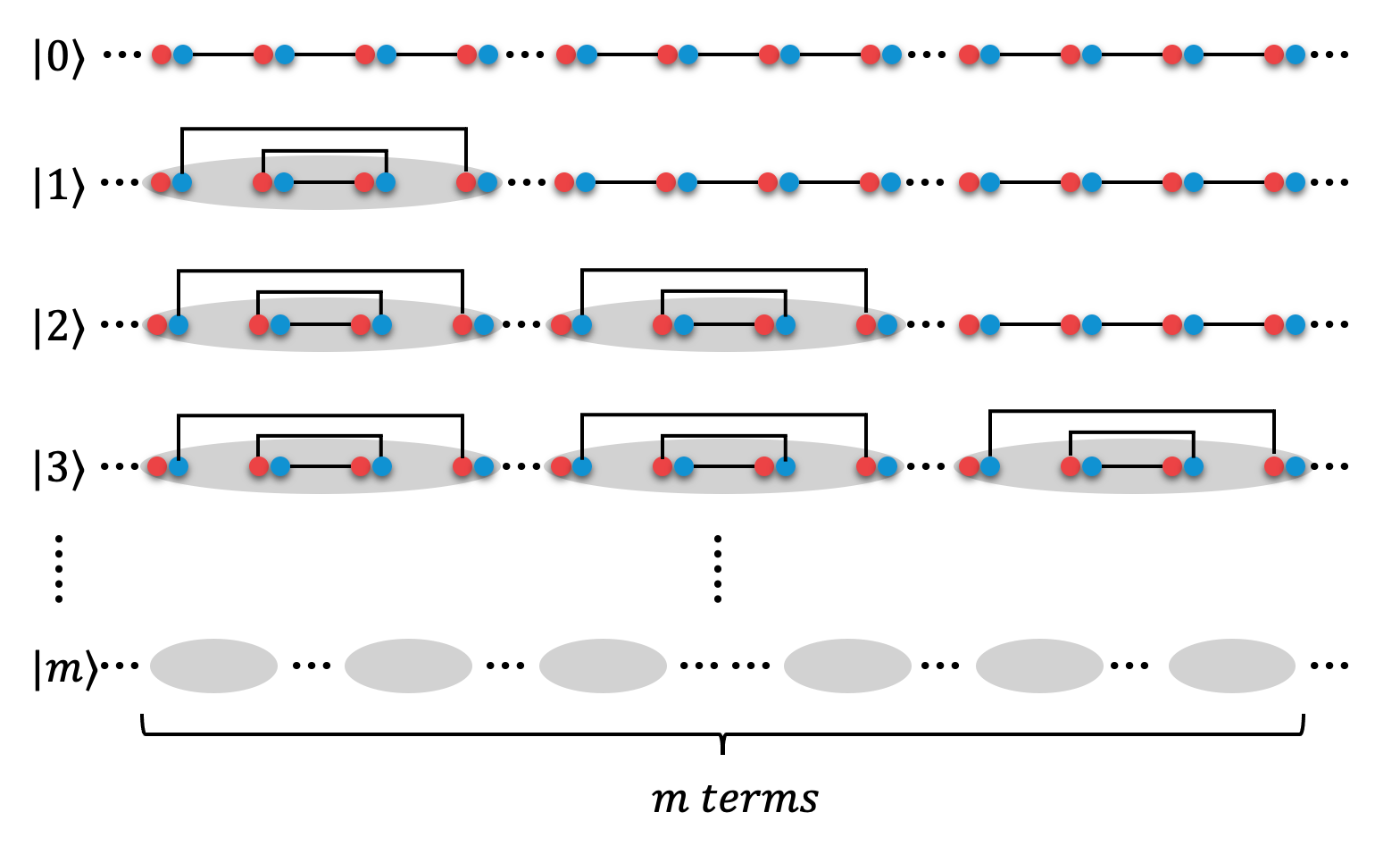}
    \caption{Illustration of different orders of excited states $\vert m\rangle$ with $m$ the number of two-bond excitations, starting from a right chiral VBS state. The $1$-st order excited state $\vert 1\rangle$ is created by breaking two right chiral singlet bonds and forming two singlets. As the order increases, more and more two-bond excitations are created. In our free bond excitations approximation, we assume that these excitations are dilute and non-interacting. Note that in this illustration we only start from the right chiral state, the same picture also happens in the left chiral configuration, which is also included in our calculations.}
    \label{fig:S5}
\end{figure}

We start with the $0$-th order state which is the VBS state of left and right chiral as
\begin{align}
\vert 0\rangle=\left(\prod_i R_i^{i+1}+\prod_i L_i^{i+1}\right)\vert vac\rangle,
\end{align}
where $R_i^{i+1}=\sum_{\sigma}a_{i,\sigma}^\dagger b_{i+1,-\sigma}^\dagger/\sqrt{3}$ and $L_i^{i+1}=\sum_{\sigma}b_{i,-\sigma}^\dagger a_{i+1,\sigma}^\dagger/\sqrt{3}$ are the right and left chiral singlet bond creation operator.
Considering the dimer Hamiltonian $H_d$ acting on this $0$-th order state, it will create bunches of two-bond excited states as
\begin{align}
\vert 1\rangle=\sum_{i}&\left(L_i^{i+1}R_{i-1}^{i+2}\prod_{j\neq i\pm 1}R_j^{j+1}+R_i^{i+1}L_{i-1}^{i+2}\prod_{j\neq i\pm 1}L_j^{j+1}\right)\vert vac\rangle,
\end{align}
and from this state the dimer Hamiltonian can again create an extra excitation to double two-bond excited state $\vert 2\rangle$. More and more excitations can be excited to the system, and we can have different numbers of two-bond excitation states $\vert n\rangle$ where $n$ labels the number of two-bond excitations.

We illustrate these states excited from the right chiral VBS one in Fig.~\ref{fig:S5}, where we have made an approximation that these two-bond excitations are far from each other, and the interaction between these excitations is negligible. This approximation is reasonable as long as there is a finite number of excitations that contribute to the system in the thermodynamic limit.

We consider a cutoff $m$ for the excitation number, and the state of the system can be written as a sum of these different excitations
\begin{align}
\vert\psi\rangle=\sum_{n=0}^mc_n\vert n\rangle,
\end{align}
where $c_n$ is the coefficient of state with free excitation number $n$. Considering the dimer Hamiltonian $H_d$ acting on these states will give
\begin{equation}
\left\{
  \begin{split}
    H_d\vert 0\rangle =& -3L\vert 0\rangle-\vert 1\rangle\\
    H_d\vert 1\rangle =& -3(L-1)\vert 1\rangle-2L\vert 0\rangle-\vert 2\rangle\\
    H_d\vert 2\rangle =& -3(L-2)\vert 2\rangle-4(L-1)\vert 1\rangle-\vert 3\rangle\\
    \vdots\\
    H_d\vert m\rangle =& -3(L-m)\vert m\rangle-2m(L-m+1)\vert m-1\rangle-\vert m+1\rangle
  \end{split}\right.,
\end{equation}
where $L$ is the length of the chain. Given the Schrodinger equation $H_d\vert\psi\rangle=E\vert\psi\rangle$ and comparing the coefficients of each state, we arrive at the following iterative equations
\begin{equation}
\left\{
  \begin{split}
    EC_0=&-3LC_0-2LC_1\\
    EC_1=&-3(L-1)C_1-4(L-1)C_2-C_0\\
    EC_2=&-3(L-2)C_2-6(L-2)C_3-C_1\\
    \vdots\\
    EC_m=&-3(L-m)C_m-2(m+1)(L-m)C_{m+1}-C_{m-1}
  \end{split}\right..
\end{equation}

\begin{figure}[h]
    \centering
    \includegraphics[width=0.5\textwidth]{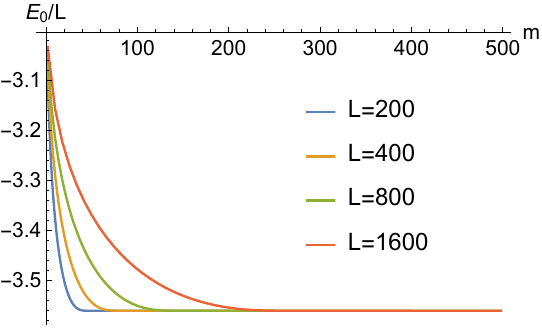}
    \caption{The ground state energy $E_0/L$ of the dimer point as a function of cut-off excitation number $m$ at the limit of free bond excitations. The energy converges to $E_0/L\approx -3.56155$ and provides an upper limit for the ground state energy of the dimer point.}
    \label{fig:S6}
\end{figure}

The above equations can be calculated iteratively by starting with $C_0=1$, and the ground state energy solution $E_0$ can be found by solving the cutoff equation $C_{m+1}=0$. We show in Fig.~\ref{fig:S6} the solution of ground state energy per site $E_0/L$ as a function of cutoff excitation number $m$ for system length $L=200, 400, 800, 1600$. As illustrated, the convergent energy is well described by cutoff excitation number $m\ll L$, which reflects the fact that the two-bond excitations are in fact dilute, and thus it is reasonable to assume that these excitations are non-interacting. Finally, we find the energy converges to $E_0/L\approx -3.56155$, which is very close to our DMRG results with $E_0/L\approx -3.594$, and provides an upper limit for this dimer point.

\begin{list}{\thesuppref}{\usecounter{suppref}}

\item\label{Cohen} M. Cohen. and T. Feldmann, J. Phys. B: At. Mol. Phys. {\bf 15}, 2563 (1982).

\item\label{Pollmann} F. Pollmann, A. M. Turner, E. Berg, and M. Oshikawa, Phys. Rev. B {\bf 81}, 064439 (2010).

\end{list}

\end{document}